\documentclass[longauth, traditabstract]{aa} 
\usepackage{graphicx}
\usepackage{txfonts}
\begin{document}
%
   \title{Strong absorption by interstellar hydrogen fluoride: 
	Herschel/HIFI observations of the sight-line to G10.6--0.4 (W31C)\thanks{Herschel is an ESA space observatory with science instruments provided
by European-led Principal Investigator consortia and with important participation from NASA.}}

   \author{D. A. Neufeld\inst{1}, P.~Sonnentrucker\inst{1}, T.~G.~Phillips \inst{2}, D.~C.~Lis \inst{2},  M.~De~Luca \inst{3}, J.~R.~Goicoechea \inst{4}, J.~H.~Black \inst{5}, M.~Gerin \inst{3}, T.~Bell\inst{2},  F.~Boulanger \inst{6}, J.~Cernicharo \inst{4}, A.~Coutens\inst{7}, E.~Dartois \inst{6}, M.~Kazmierczak\inst{8}, P.~Encrenaz \inst{3}, E.~Falgarone \inst{3}, T.~R.~Geballe \inst{9}, T.~Giesen\inst{10}, B.~Godard\inst{3},  P.~F.~Goldsmith \inst{11}, C.~Gry \inst{12}, H.~Gupta\inst{11}, P.~Hennebelle \inst{3}, E.~Herbst \inst{13}, P.~Hily-Blant\inst{14}, C.~Joblin\inst{7}, R.~Ko{\l}os \inst{15}, J.~Kre{\l}owski\inst{8}, J.~Mart\'in-Pintado\inst{4}, K.~M. Menten\inst{16}, R.~Monje\inst{2}, B.~Mookerjea \inst{17},  J.~Pearson \inst{11}, M.~Perault \inst{3}, C.~Persson \inst{5}, R.~Plume \inst{18}, M.~Salez \inst{3}, S.~Schlemmer\inst{10}, M.~Schmidt\inst{19}, J.~Stutzki \inst{10}, D.~Teyssier\inst{20}, C.~Vastel \inst{7},
S.~Yu \inst{11}, P.~Cais\inst{21}, E.~Caux\inst{7}, R.~Liseau \inst{5}, P.~Morris \inst{11},
\and P.~Planesas \inst{22}}


   \institute{The Johns Hopkins University, Baltimore, MD 21218, USA \\
 \email{neufeld@pha.jhu.edu}
\and California Institute of Technology, Pasadena, CA 91125, USA
\and LERMA, CNRS, Observatoire de Paris and ENS, France
\and Centro de Astrobiolog\`{\i}a, CSIC-INTA, 28850, Madrid, Spain
\and Chalmers University of Technology, G\"oteborg, Sweden
\and Institut d'Astrophysique Spatiale (IAS), Orsay, France
\and Universit\'e Toulouse; UPS ; CESR ; and CNRS ; UMR5187,
9 avenue du colonel Roche, F-31028 Toulouse cedex 4, France 
\and Nicolaus Copernicus University, Torun, Poland
\and Gemini telescope, Hilo, Hawaii, USA
\and I. Physikalisches Institut, University of Cologne, Germany
\and JPL, California Institute of Technology, Pasadena, USA 
\and LAM, OAMP, Universit\'e Aix-Marseille \& CNRS, Marseille, France
\and Depts.\ of Physics, Astronomy \& Chemistry, Ohio State Univ.\, USA
\and Laboratoire d'Astrophysique de Grenoble, France
\and Institute of Physical Chemistry, PAS, Warsaw, Poland
\and MPI f\"ur Radioastronomie, Bonn, Germany
\and Tata Institute of Fundamental Research, Homi Bhabha Road, Mumbai 400005, India
\and Dept. of Physics \& Astronomy, University of Calgary, Canada
\and Nicolaus Copernicus Astronomical Center, Poland
\and European Space Astronomy Centre, ESA, Madrid, Spain
\and Institute Universit\'e de Bordeaux \& CNRS, Bordeaux, France 
\and Observatorio Astron.\ Nacional (IGN) and ALMA, Santiago, Chile }
 
  \abstract
{We report the detection of strong absorption by interstellar hydrogen fluoride along the sight-line to the submillimeter continuum source G10.6--0.4 (W31C).  We have used Herschel's HIFI instrument, in dual beam switch mode, to observe the 1232.4763~GHz $J=1-0$ HF transition in the upper sideband of the Band 5a receiver.  The resultant spectrum shows weak HF emission from G10.6--0.4 at LSR velocities in the range --10 to --3 km s$^{-1}$, accompanied by strong absorption by foreground material at LSR velocities in the range 15 to 50 km~s$^{-1}$.  The spectrum is similar to that of the 1113.3430~GHz $1_{11}-0_{00}$ transition of para-water, although at some frequencies the HF (hydrogen fluoride) optical depth clearly exceeds that of para-H$_2$O.  The optically-thick HF absorption that we have observed places a conservative lower limit of $1.6 \times 10^{14}\, \rm cm^{-2}$ on the HF column density along the sight-line to G10.6--0.4.  Our lower limit on the HF abundance,  $6 \times 10^{-9}$ relative to hydrogen nuclei, implies that hydrogen fluoride accounts for between $\sim 30\%$ and 100$\%$ of the fluorine nuclei in the gas phase along this sight-line.  This observation corroborates theoretical predictions that --  because the unique thermochemistry of fluorine permits the exothermic reaction of F atoms with molecular hydrogen -- HF will be the dominant reservoir of interstellar fluorine under a wide range of conditions.}

   \titlerunning{Absorption by interstellar hydrogen fluoride}
	\authorrunning{Neufeld et al.}
   \maketitle
%

\section{Introduction}

Fluorine is alone among the elements of the periodic table in having a neutral diatomic hydride with a larger binding energy than that of molecular hydrogen.  As a result, fluorine atoms are unique in reacting exothermically with the dominant constituent of interstellar molecular clouds, H$_2$, producing hydrogen fluoride via the reaction\footnote{This reaction is arguably the most extensively-studied bimolecular chemical reaction, and has been the subject of numerous experimental and theoretical investigations (Zhu et al.\ 2002 and references therein).  Although it possesses an activation energy barrier of order $500$~K, it is believed to be moderately rapid even at low temperatures as a result of quantum mechanical tunneling.}
 $$\rm F + H_2 \rightarrow HF + H. \eqno(1)$$
Theoretical models for the interstellar chemistry of fluorine-bearing molecules (Neufeld, Wolfire \& Schilke 2005; Neufeld \& Wolfire 2009) have made the prediction that hydrogen fluoride will be the dominant reservoir of gas-phase fluorine over a wide range of conditions as a direct result of reaction (1).  In particular, HF (hydrogen fluoride) is expected to be abundant even close to molecular cloud surfaces that are irradiated by the interstellar ultraviolet radiation field; indeed, HF will be the dominant fluorine-bearing species wherever hydrogen is predominantly molecular.  Thus, these models make the surprising prediction that HF can be the most abundant molecule after H$_2$
close to UV-irradiated cloud surfaces; here the expected abundance of HF exceeds that of CO despite the fact that F nuclei are less abundant than C nuclei by more than four orders of magnitude (Asplund et al.\ 2009).  Indirect support for this prediction has come from the detection of the CF$^+$ molecular ion, by means of ground-based observations of its $J=1-0$, $J=2-1$ and $J=3-2$ transitions towards the Orion bar (Neufeld et al.\ 2006) and of its $J=1-0$ transition towards several other UV-irradiated molecular clouds (Neufeld et al.\ 2010, in preparation).  Because the presumptive production mechanism for CF$^+$ is the reaction
$$\rm HF + C^+ \rightarrow CF^+ + H, \eqno(2)$$ these detections of CF$^+$ argue for a significant overlap of the regions where HF and C$^+$ are both abundant, and thus for the presence of large HF abundances close to cloud surfaces where the C$^+$/CO ratio is large. 

The detection of interstellar HF itself, however, is severely hampered by atmospheric absorption.  Because HF has the smallest moment of inertia of any molecule containing a heavy element, its pure rotational transitions lie at high frequencies inaccessible from ground-based observatories.  To date, the only detection of interstellar HF has been obtained toward a single source by means of observations (Neufeld et al.\ 1997) performed with the {\it Infrared Space Observatory (ISO)}.  In that study, the $J=2-1$ transition at 2.463~THz was detected in absorption toward the strong submillimeter continuum source Sgr B2.  Because the spectral coverage of {\it ISO} did not extend to the 1.232~THz $J=1-0$ transition, the search for HF in absorption was impossible except in exceptional regions where the $J=1$ state was significantly populated by radiative excitation.

The Heterodyne Instrument for the Far-Infrared (HIFI\footnote{HIFI has been designed and built by a consortium of institutes and university departments from across
Europe, Canada and the United States under the leadership of SRON Netherlands Institute for Space
Research, Groningen, The Netherlands and with major contributions from Germany, France and the US.
Consortium members are: Canada: CSA, U.~Waterloo; France: CESR, LAB, LERMA, IRAM; Germany:
KOSMA, MPIfR, MPS; Ireland, NUI Maynooth; Italy: ASI, IFSI-INAF, Osservatorio Astrofisico di Arcetri-
INAF; Netherlands: SRON, TUD; Poland: CAMK, CBK; Spain: Observatorio Astron\'omico Nacional (IGN),
Centro de Astrobiolog\'a (CSIC-INTA). Sweden: Chalmers University of Technology - MC2, RSS \& GARD;
Onsala Space Observatory; Swedish National Space Board, Stockholm University - Stockholm Observatory;
Switzerland: ETH Zurich, FHNW; USA: Caltech, JPL, NHSC.}; De Graauw et al.\ 2010), on board the {\it Herschel Space Observatory} (Pilbratt et al.\ 2010), provides access to the HF $J=1-0$ transition for the first time.  Thus, absorption line spectroscopy of bright submillimeter continuum sources, performed with HIFI, can provide a sensitive probe of HF in its {\it ground rotational state} ($J=0$).  As part of the PRISMAS (''PRobing InterStellar Molecules with Absorption line Studies'') Key Program, we will exploit this new capability by performing HIFI observations of eight strong submillimeter continuum sources with sight-lines that are known to intercept foreground molecular material.  In this {\it Letter}, we present the first results of our search for hydrogen fluoride, obtained from observations of the sight-line to G10.6--0.4 (W31C).  

G10.6--0.4 is a region of high-mass star-formation.  It is one of three bright HII regions within the W31 complex, and an extremely luminous submillimeter and infrared continuum source; Wright, Fazio \& Low (1977) estimate its infrared luminosity as $\sim 10^7 L_\odot$.  Using observations of foreground HI 21 cm absorption, Fish et al.\ (2003) have obtained a kinematic distance estimate of $4.8^{+0.4}_{-0.8}$~kpc for G10.6--0.4, a value that places the source within the so-called ``--30~km~s$^{-1}$'' spiral arm.  The sight-line to G10.6--0.4 intersects several foreground molecular clouds, the arrangement of which has been elucidated by Corbel \& Eikenberry (2004; see their Figure 8).    Because G10.6--0.4 shows a large continuum flux at infrared and radio wavelengths, this sight-line has proven one of the most valuable in the Galaxy for the study of interstellar gas by absorption line spectroscopy.  Indeed, line absorption by foreground gas has been detected over three and one-half decades in wavelength, from species as diverse as atomic oxygen (at 63~$\mu$m; Keene et al.\ 1999), $^{13}$CH$^+$ ($J=1-0$ at 361~$\mu$m; Falgarone, Phillips \& Pearson 2005), HCO$^+$ ($J=1-0$ at 3.36~mm; Keene et al.\ 1999; Godard et al.\ 2010), and atomic hydrogen (at 21~cm; Fish et al.\ 2003). 


\section{Observations and data reduction}

We observed the $J=1-0$ transition of HF, with rest frequency 1232.4763~GHz (Nolt et al.\ 1987), in the upper sideband of the Band 5a HIFI receiver.  To help determine whether any observed feature did indeed lie in the upper sideband, three observations were carried out with slightly different settings of the local oscillator (LO) frequency.  These observations, each with an on-source integration time of 75~s, were carried out on 2010 March 5, using the dual beam switch (DBS) mode and the Wide Band Spectrometer (WBS).  The WBS has a spectral resolution of 1.1~MHz, corresponding to a velocity resolution of $0.27\, \rm km\,s^{-1}$ at the frequency of the HF $J=1-0$ transition. The telescope beam, of diameter $\sim 18^{\prime\prime}$ HPBW, was centered at 
$\rm \alpha=18h\,10m\,28.7s, \delta = -19^0\, 55^\prime \,50.0^{\prime\prime} (J2000)$.  
The reference positions for these observations were located 3$^\prime$ on either side of the source along an East-West axis; by taking the difference between the reference position spectra, we confirmed that the reference positions are devoid of measurable line or continuum emission at the frequencies of present interest.

The data were reduced using the standard Herschel pipeline to Level 2, providing fully calibrated spectra of the source.  The Level 2 data were analysed further using the Herschel Interactive Processing Environment (HIPE), along with ancillary IDL routines that we have developed.  The signals measured in the two orthogonal polarizations were in excellent agreement, as were spectra obtained at the three LO settings when assigned to the upper sideband.  We combined the data from the three observations, and from both polarizations, to obtain an average spectrum. 

\begin{figure}
\includegraphics[width=9 cm]{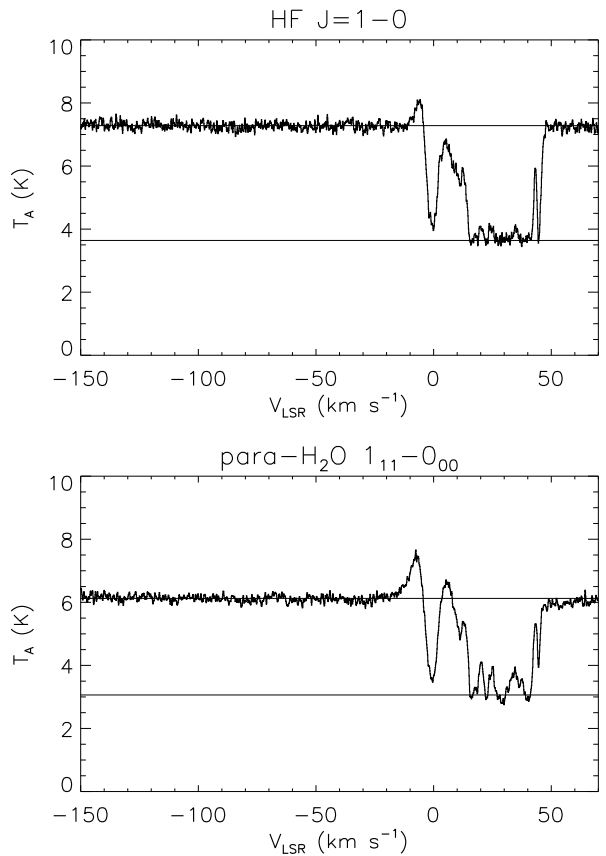}
\caption{Spectra of HF $J=1-0$ ({\it upper panel}) and para-H$_2$O $1_{11}-0_{00}$ ({\it lower panel}) obtained toward G10.6--0.4  Note that because HIFI employs double sideband receivers, the complete absorption of radiation at a single frequency will reduce the measured antenna temperature to one-half the apparent continuum level. 
}
\end{figure}

\section{Results}

The upper panel of Figure 1 shows the WBS spectrum of HF $J=1-0$, with the frequency scale expressed as Doppler velocities relative to the Local Standard of Rest (LSR).  The data quality is excellent, with no significant baseline ripples.  The double sideband continuum antenna temperature is $T_A({\rm cont}) = 7.28$~K, and the r.m.s noise is 0.12~K.  Because HIFI employs double sideband receivers, the complete absorption of radiation at a single frequency will reduce the measured antenna temperature to one-half the apparent continuum level (given a sideband gain ratio of unity).  Horizontal lines in Figure 1 indicate the values of $T_A({\rm cont})$ and $0.5\,T_A({\rm cont})$.  Thus, the spectrum plotted in the upper panel of Figure 1 is consistent with a sideband gain ratio close to unity and a large optical depth for HF absorption over the entire range $15 < {\rm v}_{\rm LSR} < 50\, \rm km \, s^{-1}$.  In addition, weak HF emission is clearly evident for v$_{\rm LSR}$ in the range --10 to --3 $\rm km \, s^{-1}$.

\begin{figure}
\includegraphics[width=9 cm]{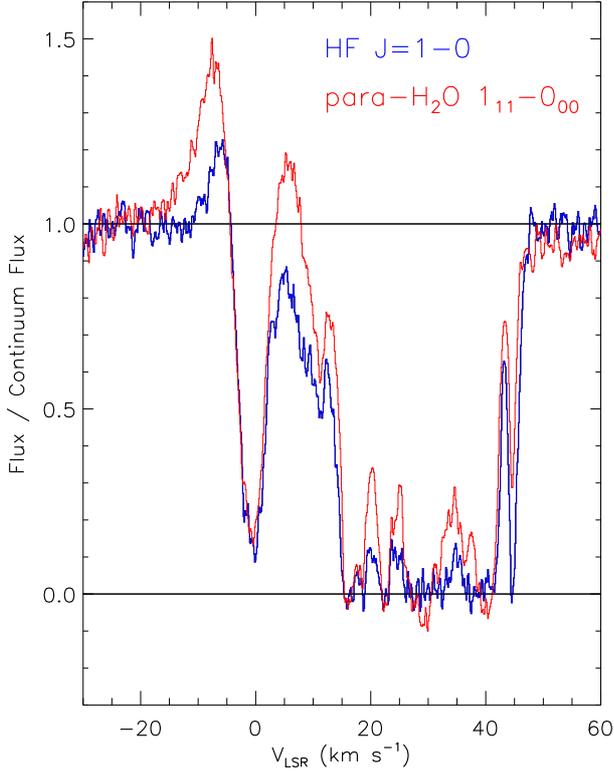}
\caption{Spectra of HF $J=1-0$ (blue) and para-H$_2$O $1_{11}-0_{00}$ (red) over the LSR velocity range --30 to 60 $\rm km\, s^{-1}$.}
\end{figure}

In the lower panel of Figure 1, we present for comparison the spectrum of the $1_{11} - 0_{00}$ 1113.3430~GHz line of para-water, obtained by HIFI on 2010 March 5 in the lower sideband of the same Band 5a receiver.  The methods used to acquire and reduce the para-water data were entirely analogous to those described for HF $J=1-0$ in the preceding section, the only difference being that on-source integration time for each observation was 39~s instead of 75~s.
In Figure 2, we present the same data over a narrower range of v$_{\rm LSR}$.  In this figure, we present the flux, normalized with respect to the continuum flux, in a single sideband.  Here, we assumed a sideband gain ratio of unity, so we plot the quantity $2T_A/T_A({\rm cont}) - 1$. 

\begin{figure}
\includegraphics[width=9 cm]{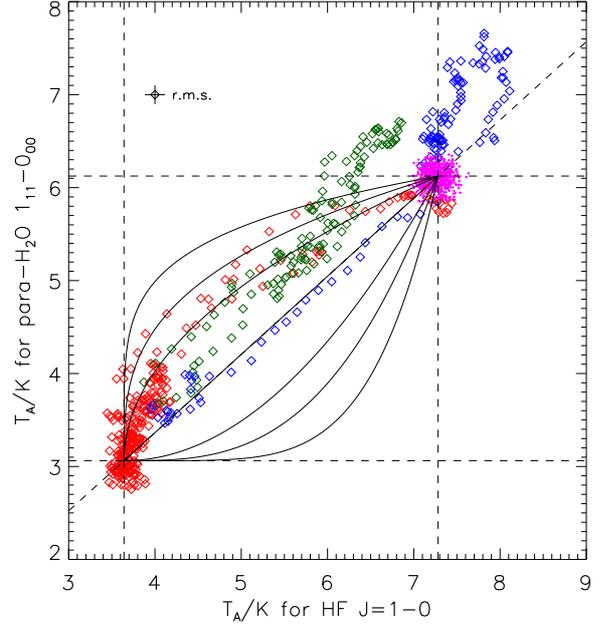}
\caption{HF $J=1-0$ antenna temperature versus para-H$_2$O $1_{11}-0_{00}$ antenna temperature (see text for details)}
\end{figure}

\section{Discussion}

The spectra shown in Figure 2 can be divided into two regions.  Close to the systemic velocity of G10.6--0.4, determined from observations of OH maser emission to lie at v$_{\rm LSR}=-1 \, \rm km\,s^{-1}$, we observe a self-absorbed emission/absorption feature associated with the source itself, while at LSR velocities in the range $\sim 15 \rm \, km \,s^{-1}$ to $\sim 50 \rm \, km \,s^{-1}$
we observe absorption by foreground material unassociated with G10.6--0.4.  This range of LSR velocities is characteristic of absorption by other species along this sight-line (Godard et al.\ 2010), including HCO$^+$ (Keene et al.\ 1999) and atomic hydrogen (Fish et al.\ 2003). 
In this {\it Letter}, we focus on the absorption by foreground material; the self-absorbed emission/absorption features associated with G10.6--0.4 will be the subject of a future paper.

The similarity between the HF and para-H$_2$O spectra is striking, and remarkable given the fact that F nuclei are less abundant than O nuclei by more than 4 orders of magnitude.  At some frequencies, the HF transition clearly shows a larger optical depth than the para-water transition, particularly near LSR velocities of 20, 25, and 45 $\rm \, km \, s^{-1}$.  In Figure 3, we plot the values of the antenna temperatures for the two spectral lines, with each point representing a single velocity channel.  Magenta, blue, green, and red points apply respectively to the LSR velocity ranges --150 to --15 $\rm km \,s^{-1}$ (continuum emission); --15 to 0 $\rm km \,s^{-1}$ (emission and absorption associated with G10.6--0.4); 0 to 15 $\rm km \,s^{-1}$ (material associated with G10.6--0.4, but possibly affected by foreground absorption); and 15 to 50 $\rm km \,s^{-1}$ (foreground absorption).  In this plot, dashed horizontal and vertical lines indicate values of $T_A({\rm cont})$ and $0.5 \, T_A({\rm cont})$.  Solid black lines indicate the loci expected for given ratios of the HF to H$_2$O optical depth; from top to bottom, these ratios are 5, 3, 2, 1, 1/2, 1/3, and 1/5.  Clearly, when the optical depths are either large or small, this ratio is poorly constrained.  However, in several channels where the spectrum is dominated by foreground absorption (red points) and the optical depths are moderate, it is clear the HF optical depth can be as large as $\sim 5$ times the H$_2$O optical depth.  

Because of their large spontaneous radiative decay rates ($2.42 \times 10^{-2}\,\rm s^{-1}$ and $1.84 \times 10^{-2}\,\rm s^{-1}$ respectively for the HF and para-H$_2$O transitions), both 
the transitions we have observed possess high critical densities\footnote{Here, ``critical density'' is defined as the gas density at which the collisional deexcitation rate equals the spontaneous radiative decay rate.}.  
Thus, except within the dense gas associated with the core of G10.6--0.4, we expect that each of the two species we have observed will be almost entirely in its ground rotational state.  In that case, the absorption optical depth for the HF $J=1-0$ transition, integrated over velocity, is given by
$$\int \tau dv = {A_{ul} g_u \lambda^3 \over 8 \pi g_l} N({\rm HF}) = 4.16 \times 10^{-13} N({\rm HF}) \, \rm cm^2 \, km \, s^{-1}, \eqno(3)$$
where $A_{ul} = 2.42 \times 10^{-2}\,\rm s^{-1}$ is the spontaneous radiative decay rate, $g_u=3$ and $g_l=1$ are the degeneracies of the upper and lower states, and $\lambda=243.2\,\mu$m is the transition wavelength.  
The analogous expression for the para-H$_2$O $1_{11}-0_{00}$ transition is 
$$\int \tau dv  = 4.30 \times 10^{-13} N({\rm p-H_2O}) \, \rm cm^2 \, km \, s^{-1}. \eqno(4)$$
Given these expressions, and an assumed ortho-to-para ratio of 3 for water, the opacity ratios indicated in Figure 3 suggest that the water and HF abundances are almost equal in clouds of moderate optical depth (but unconstrained at LSR velocities where the optical depths are either large or small).

In the region where foreground absorption dominates, the optical depth may be estimated as --ln$\,(2T_A/T_A({\rm cont})-1)$.  However, at frequencies where the optical depth is large, its exact value is highly uncertain, being affected strongly by noise and by any small differences in the gains for the two sidebands.  Accordingly, we place a conservative lower limit of ln(10) on the optical depth at frequencies for which $(2T_A/T_A({\rm cont})-1) \le 0.1$.  With this assumption, we obtain lower limits on the HF and para-H$_2$O column densities of $1.6 \times 10^{14} \, \rm cm^2$ and $1.4 \times 10^{14} \, \rm cm^2$ respectively for absorbing material with LSR velocities in the range 15 to 50 $\rm \, km \, s^{-1}$.  In the v$_{\rm LSR}$ range 0 to 15 $\rm \, km \, s^{-1}$, the spectrum is affected both by the source itself and by the foreground material; we therefore exclude this velocity range from our calculation of the foreground column densities, making our lower limits doubly conservative.

To determine the average HF abundance within the foreground absorbing material, we require estimates of the total column density along the sight-line.  Unfortunately, foreground H$_2$ is not directly detectable, and thus our estimate is necessarily indirect.  Based upon infrared stellar spectroscopy and photometry, Blum, Damineli \& Conti (2001) have estimated a K-band extinction $A_K=1.71 \pm 0.19$ along a sight-line to a star cluster associated with G10.2--0.3, an HII region located in the W31 complex approximately 30$^\prime$ southwest of G10.6--0.4. For an assumed extinction ratio $A_V/A_K \sim 8$ (Cardelli, Clayton \& Mathis 1989), this corresponds to $\sim 14$~mag of visual extinction.   This extinction estimate is broadly consistent with an independent analysis carried out by Corbel \& Eikenberry (2004).  Based on observations of CO emission along a third nearby sight-line to G10.0--0.3, Corbel \& Eikenberry identified five molecular clouds between the Sun and G10.6--0.4: these clouds, designated MC4, MC24, MC30, MC38, and MC44, lie at estimated (kinematic) distances of roughly 0.2, 3.0, 3.5, 4.2, and 4.5 kpc from the Sun.  Estimating the extinction contributed by each cloud on the basis of its integrated CO brightness temperature, Corbel \& Eikenberry obtained a total visual extinction $A_v \sim 13$~mag for the set of foreground clouds listed above.  Although obtained along two different sight-lines, both of which are offset from the one that we have observed, we adopt these extinction estimates for G10.6--0.4.  Given an $N_H / A_V$ ratio of $1.9 \times 10^{21} \rm \, cm^{-2}$ (Bohlin, Savage \& Drake 1978), this extinction estimate implies a value $\sim 2.7 \times 10^{22}\, \rm \, cm^{-2}$ for the total column density of H nuclei, $N_H = N({\rm H}) + 2N({\rm H_2})$, along the sight-line to G10.6--0.4.  This estimate is a factor $\sim 4$ larger than that adopted by Falgarone et al.\ in their recent study of $^{13}$CH$^+$ absorption toward G10.6--0.4.  The reason for this discrepancy is that Falgarone et al.\ were concerned with the column density of material possessing a molecular to atomic hydrogen ratio $\le 0.5$; in the present analysis, we seek to determine HF and para--H$_2$O abundances averaged over all the material along the line-of-sight.  Our estimate of $N_H$ is more than twice the atomic hydrogen column density inferred by Godard et al.\ (2010) from the 21~cm absorption observations of Fish et al., implying that the absorbing material is predominantly molecular.   

Given the estimate $N_H = 2.7 \times 10^{22}\, \rm \, cm^{-2}$ for the total column density of H nuclei, we obtain a conservative lower limit $N({\rm HF})/N_H \ge 6 \times 10^{-9}$.  By comparison, the solar and meteoritic elemental abundances for fluorine are $N_F/N_H = 3.6^{+3.6}_{-1.8} \times 10^{-8}$ and $2.63^{+0.39}_{-0.34} \times 10^{-8}$, respectively (Asplund et al.\ 2009), whilst the average gas-phase interstellar abundance in diffuse atomic gas clouds 
is $N_F/N_H = 1.8 \times 10^{-8}$ (Snow, Destree \& Jensen \ 2007).  Thus, our observations suggest that HF accounts for at least one-third of fluorine in the gas phase along the sight-line to G10.6--0.4.

\section{Conclusions}

Our detection of optically-thick absorption in the HF $J=1-0$ transition over a broad range of LSR velocities corroborates the theoretical prediction that HF is the dominant reservoir of gas-phase fluorine under a wide variety of interstellar conditions.  This conclusion suggests that HIFI observations of HF will provide a useful probe of diffuse molecular gas throughout the Galaxy, and -- if its high abundance is confirmed along other similar sight-lines for which upcoming HIFI observations are planned -- a potentially valuable surrogate for molecular hydrogen. 
Along sight-lines where absorbing material covers a substantial range of LSR velocities, HF absorption should be detectable (although spectrally unresolved) even at the resolution of Herschel's SPIRE instrument (Griffin et al.\ 2010); this opens up the possibility of detecting hydrogen fluoride toward suitable extragalactic sources as well.

\begin{acknowledgements}
This research was performed through a JPL contract funded by the National Aeronautics and Space Administration.
\end{acknowledgements}

\end{document}